\documentclass[epj]{svjour}

\usepackage{graphicx} % Required for the inclusion of images
\usepackage{subfigure} % use subfigures in the figure environment
\usepackage{amsmath} % Required for some math elements 
\usepackage{float} % used to fix the position of a figure
\usepackage{epsfig}
\usepackage{epstopdf}

\usepackage{graphics}
\usepackage{graphicx}
\usepackage{grffile} %deal with fig name with ".,_ " .etc...

\usepackage{amsmath}

\usepackage{color}

\usepackage{hyperref}

\newcommand{\snn}{\sqrt{s_{NN}}}
\begin{document}

%moved in document following aps style
\title{Investigating the NCQ scaling of elliptic flow at LHC with a multiphase transport model}

\author{Liang Zheng\inst{1}\inst{2}\thanks{zhengl@mails.ccnu.edu.com} \and Hui Li\inst{1}
\and Hong Qin\inst{2} \and Qi-Ye Shou\inst{1}\thanks{qiye.shou@cern.ch} \and Zhong-Bao Yin\inst{1}\thanks{zbyin@mail.ccnu.edu.cn}}
\institute{Key Laboratory of Quark and Lepton Physics (MOE) and Institute of Particle Physics, Central China Normal University, Wuhan 430079, China,
\and School of Mathematics and Statistics, Central China Normal University, Wuhan, 430079, China}

%\author{Hui Li}
%\institute{Key Laboratory of Quark and Lepton Physics (MOE) and Institute of Particle Physics, Central China Normal University, Wuhan 430079, China}
%\author{Hong Qin}
%\institute{Department of Mathematics and Statistics, Central China Normal University, Wuhan, 430079, China}
%\author{Qi-Ye Shou} \thanks{qiye.shou@cern.ch}
%\institute{Key Laboratory of Quark and Lepton Physics (MOE) and Institute of Particle Physics, Central China Normal University, Wuhan 430079, China}
%\author{Zhong-Bao Yin}
%\institute{Key Laboratory of Quark and Lepton Physics (MOE) and Institute of Particle Physics, Central China Normal University, Wuhan 430079, China}
%

\date{\\}%Received: \today / Revised version: \today}

\abstract{
The number of constituent quark (NCQ) scaling behavior of elliptic flow has been
systematically studied at the LHC energy within the framework of a multiphase transport
model (AMPT) in this work. With the variation of the fragmentation parameters, collision
centrality and system energy, we find that the initial conditions of parton dynamics are
more important than the final state parton cascade process for the existence of NCQ
scaling when the hadronic interaction is off in Pb-Pb collisions. By turning on the hadron
interaction process, the impacts of hadronic evolution are found to be responsible for a
significant violation to the well established scaling structure. Our study suggests the
interpretation of NCQ scaling is not only subject to the hadronization mechanism but also
to the initial conditions of parton evolution as well as the hadronic interactions
especially for the LHC experiments.
}

 % end of PACS codes
     
%\authorrunning{L. Zheng and Z.B. Yin}
%\titlerunning{Initial state of heavy ion collisions with quark participant assumption}

%comes after abstract following aps style
\maketitle

\section{Introduction}

The formation and the evolution of the color-deconfined strongly interacting matter, quark-gluon plasma (QGP)~\cite{Gyulassy:2004zy}, have been studied via relativistic heavy-ion collisions for decades. As one of the most important quantities, elliptic flow ($v_2$), which is defined as the second-order Fourier component of the particle azimuthal distribution~\cite{Kolb:2003dz}, receives considerable attention and plays an essential role in studying the collective motion and bulk property of the QGP. The transverse momentum ($p_T$) dependence of $v_2$ serves as an unique probe to test different physics processes. At low $p_T$, $v_2$ is commonly used to extract the information of the early stage of the collisions, such as the pressure gradients, the degree of thermalization, and the equation of state, {\it etc}~\cite{Shen:2011eg,Gyulassy:2004zy,Adamczyk:2013gw}. In addition, for a given low $p_T$ value, $v_2$ of different particles are observed to be ordered by mass with heavier particles having the smaller $v_2$ and vice versa.

In the intermediate $p_T$ region, $v_2$ is considered to be able to reveal the production mechanism of hadron. Results from RHIC show the $p_T$ differential $v_2$ of particles tend to group based on their hadron types, baryons or mesons~\cite{Abelev:2007qg,Adler:2003kt,Adams:2004bi}. This phenomenon has been interpreted as a signature of quark recombination or coalescence models, in which hadrons are formed from the combination of the neighboring constituent quarks in phase space~\cite{Molnar:2003ff,Jia:2006vj,Han:2011iy}. 
More intriguingly, it is also proposed by such models that $v_2$ of different hadrons, when scaled by the number of constituent quarks (NCQ or $n_q$), could approximately exhibit a uniform behavior with $v_2$/$n_q$ vs. $p_T$/$n_q$ falling on a universal curve, namely,  
\begin{equation}
\frac{v_{2}^{B}(p_{T}/3)}{3} = \frac{v_{2}^{M}(p_{T}/2)}{2},
\end{equation}
where superscript $B$ ($M$) denotes baryons (mesons) consisting of 3(2) constituent quarks.

This feature has been observed at RHIC with surprising accuracy~\cite{Afanasiev:2007tv,Adams:2003am,Adare:2006ti,Adamczyk:2015fum,Abelev:2008ae}, and furthermore, it can also manifest itself when investigating the transverse kinetic energy ($m_{T}-m_{0}$) dependence of $v_2$~\cite{Afanasiev:2007tv,Tang:2011xq}. Such a scaling can be understood as a consequence of the coalescence mechanism and the deduction of $v_2$ in the partonic evolution stage. For a narrow hadronic wave function in momentum space, the $v_2$ of a given type of meson with valence quark $a$ and $b$ reads~\cite{Fries:2008hs},
\begin{equation}
v_{2}^{M}(p_{T}) = v_{2}^{a}(x_{a}p_{T}) + v_{2}^{b}(x_{b}p_{T}),
\label{eq:mab}
\end{equation}
where $x_{a}$ and $x_{b}$ are fixed momentum fractions with $x_{a}$ + $x_{b}$ = 1. In the case of $a$ and $b$ having the same $v_2$, Eq. (\ref{eq:mab}) can be simplified to
\begin{equation}
v_{2}^{h}(p_{T}^{h}) = n_{q}v_{2}^{q}(p_{T}^{h}/n_{q}),
\end{equation}
where superscript $h$ ($q$) represents hadrons (quarks) and $n_{q}$ = 3 (2) for baryons (mesons). This treatment has also been extended to higher-order flow harmonics~\cite{Sun:2014dda,Zhang:2015skc}.

For many years, it is widely believed that the NCQ scaling is a natural result and an important evidence of the dominance of quark degrees of freedom in the evolution of QGP. Recently, however, both PHENIX~\cite{Adare:2012vq} and ALICE experiments~\cite{Abelev:2014pua} have reported the deviation of the scaling properties in heavy-ion collisions, and the latter, in particular, announced the violation at the level of $\pm$20$\%$ for all centrality intervals, which triggers various theoretical concerns and reconsiderations~\cite{Tian:2009wg,Dunlop:2011cf,Lu:2006qn}. In Ref.~\cite{Singha:2016aim}, the authors argue that the high phase-space density of constituent quarks should be the key reason of the broken of NCQ scaling. Nevertheless, the formation of the scaling properties is the consequence of a succession of interaction processes at both partonic-level and hadronic-level. Therefore, it is worthwhile to study the integrated effect instead of a single factor through a transport model to understand the contribution from each intermediate process.

In this work we perform a systematic study of the possible factors contributing to the deviation of the $v_2$ NCQ scaling at the LHC energy with the string-melting version of the AMPT model, in which the parton coalescence mechanism and hadron cascade process are jointly provided. This paper is organized as follows. In Sec.~\ref{sec:ampt} and Sec.~\ref{sec:epmethod}, we give a brief introduction to the AMPT model and the methodology to calculate the azimuthal flow in our study. The results of identified particle flow confronted with NCQ scaling is presented in Sec.~\ref{subsec:results_ncq_flow}. To have a systematic understanding to the collective anisotropy at the LHC, we also study the centrality dependence of the constituent quark number scaled flow behavior in Sec.~\ref{subsec:results_sys}. We further explore the hadronic evolution effects including resonance decay and hadron rescattering for different particles in Sec.~\ref{subsec:results_hadeff}. The energy dependence of the NCQ scaling has been studied by investigating identified flow at $\snn=$ 5.02 TeV in Sec.~\ref{subsec:results_prediction}.

\section{AMPT model}
\label{sec:ampt}
The AMPT model~\cite{Lin:2004en} is a hybrid transport model widely used in the
description of collective behavior in heavy ion collisions. It uses initial
parton conditions generated in HIJING~\cite{Wang:1991hta}. The parameters $a$ and $b$ used
in HIJING determine the Lund string fragmentation function as $f(z)\propto
z^{-1}(1-z)^{a}\exp{(-bm^{2}_{\perp}/z)}$, where $z$ is the light-cone momentum
fraction of the produced hadron of transverse mass $m_{\perp}$ with respect to
the fragmenting string. 

In the string melting version of AMPT model, initial strings from HIJING are
converted to their valence quarks and antiquarks. The evolution of the partonic
phase is implemented with the Zhang's Parton Cascade (ZPC)
model~\cite{Zhang:1997ej}, which calculates the parton-parton scattering using
cross sections $\sigma\approx\frac{9\pi\alpha_{s}^{2}}{2\mu^{2}}$ based on a
Debye screening mass $\mu$. After the partons stop interaction in ZPC, a
hadronization process based on the quark coalescence model has been applied to
the quarks, which combines the nearest quarks in coordinate space into hadrons.
Hadronic stage evolution is then performed to the hadrons formed during quark
coalescence, handled by the ART model with the input cross section for different
hadron-hadron scattering channels. The hadronic interactions in this step will
be terminated at a cutoff time $t_{max}=30$ fm/$c$ by default, when all the
observables of interest are stable at final state.

In this work we use the program version with string melting
Ampt-v1.26t5-v2.26t5. It is proposed that in this version of AMPT model the set
of Lund string fragmentation parameters $a=0.3$, $b=0.15$ GeV$^{-2}$, (set A) along with
$\sigma=3$ mb and $\mu=2.3$ fm$^{-1}$ can be used to describe the rapidity density,
transverse momentum spectrum and elliptic flow at the LHC energy
simultaneously~\cite{Lin:2014tya}. Compared to other widely used parameter sets,
a small $b$ is applied to generate large mean transverse momentum of the initial
hadrons in this tuning. As the effective string tension $\kappa\propto
1/[b(2+a)]$ is quite large due to the selection of $b$ value, an upper limit of
PARJ(2) on the relative production of strange to nonstrange quarks is set to 0.4
to constrain the strangeness production. We will stick to this parameter set in the
following study, unless otherwise specified.

\section{Event Plane Method}
\label{sec:epmethod}
In this study, we will use the event plane method as illustrated
in~\cite{Voloshin:2008dg} to calculate the elliptic flow in AMPT. The procedures
of this method will be briefly described here. The momentum distribution of
emitted particles can be expressed as follows:
\begin{equation}
E\frac{d^3N}{d^3p}=\frac{1}{2\pi}\frac{d^2N}{p_{T}dp_{T}dy}(1+\sum_{n} 2v_{n}\cos[n(\phi-\Psi_n)]),
\end{equation}
where $\phi$ is the azimuthal angle of the particle in transverse plane,
$y$ and $p_{T}$ represent the rapidity and transverse momentum,
$v_{n}$ is the $n$-th harmonic flow. The event plane angle $\Psi_{n}$ is
determined with the momentum of all the particles within $|\eta|<0.8$:
\begin{equation}
\Psi_{n}=\frac{1}{n}[\arctan{\frac{\sum_{i} p_{T}^{i}\sin(n\phi_{i})}{\sum_{i} p_{T}^{i}\cos(n\phi_{i})}}],
\end{equation}
where the subscript $i$ runs over all the selected particles in that event.
The selected particle of interest has been excluded when building up
the corresponding event plane to avoid the auto-correlation.
The azimuthal flow with respect to this event plane is:
\begin{equation}
v_{n}(p_{T})=\langle \cos{[n(\phi-\Psi_n)]} \rangle. \label{eqn:flow}
\end{equation}
The bracket $\langle \cdots \rangle$ denotes an average over all particles in
the event sample with transverse momentum $p_T$. Since the reconstructed event
plane is only an approximation to the true reaction plane, the observed
azimuthal flow coefficient obtained in Eq.~\ref{eqn:flow} has to be corrected by
the event plane resolution. In this work, we restrict ourselves to the elliptic
flow of particles within pseudorapidity window $|\eta|<0.8$ for Pb-Pb collisions
at the LHC energy. Meanwhile, due to the large multiplicity generated at the LHC
energy scale, the resolution of $\Psi_{n}$ can be practically taken to be 100\%
for central and semi-central collisions. 

\begin{figure}
\centering
\includegraphics[width=0.45\textwidth]{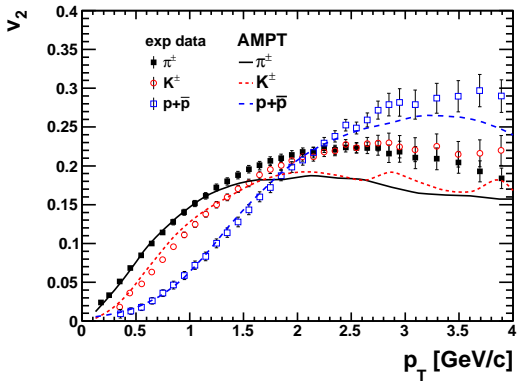}
\caption{Differential $v_2$ on $p_T$ of $\pi$ K and p in mid-rapidity for Pb-Pb collisions at $\snn=2.76$ TeV for 
centrality 30-40\% with input parameters $a=0.3$, $b=0.15$ GeV$^{-2}$, $\sigma=3$ mb and $\mu=2.3$ fm$^{-1}$.
Data is from the ALICE experiment~\cite{Abelev:2014pua}.}
\label{fig:v2_data_compare}
\end{figure}

The comparison of the AMPT model simulation results with parameters listed in
the last section and the measured experimental data from
ALICE~\cite{Abelev:2014pua} is shown in Fig.~\ref{fig:v2_data_compare} for the
identified particle elliptic flow at Pb-Pb $\snn=2.76$ TeV with centrality
30-40\%. It is observed that the AMPT model generated $v_2$ agrees very well with the
ALICE data to a large extent.

\section{Results and Discussions}
\label{sec:results}

\subsection{Identified particle elliptic flow at Pb-Pb $\snn=2.76$ TeV without hadronic evolution}
\label{subsec:results_ncq_flow}
The anisotropy in AMPT is developed initially through parton level evolution. 
The hadron level flow is derived from the partonic anisotropy
through a quark coalescence model and then modified in the hadronic evolutions.
We exclude the hadron phase contribution by focusing on the primordial particles
generated directly after coalescence procedure at this moment, and leave the
discussion of hadronic evolution effect to the later part in
Sec.~\ref{subsec:results_hadeff}. In the following text, we refer to the
particles formed right after coalescence as the primordial particles.

Another parameter set tuned to be applicable both at RHIC and LHC energy scale
with $a=0.5$, $b=0.9$ GeV$^{-2}$, (set B) $\sigma=1.5$ mb and
$\mu=3.2$ fm$^{-1}$~\cite{Xu:2011fi,Xu:2011fe} has also been used in a wide range of
phenomenological studies. The set B parameters, corresponding to a smaller 
string tension $\kappa_{B}\approx 1/6\kappa_{A}$, are found to reasonably reproduce 
the azimuthal flow but underestimates the particle yield for large $p_T$
at the LHC. 
To provide a systematic understanding on the behavior of the identified particle
elliptic flow formed in AMPT, we will compare the effects of set A and set B (details can be found int Tab.~\ref{tab:set})
with the same parton-parton cross section $\sigma=$ 3 mb. Meanwhile, a variation
on the parton scattering cross section $\sigma=$ 10 mb based on set A has also
been provided. The simulation has been performed for Pb-Pb collisions at $\snn=2.76$ TeV 
with the centrality of 30-40\%. 
\begin{table}[hbt]
\centering
\caption{Details of set A and set B parameters}
\begin{tabular}{p{3cm}|p{2cm}p{2cm}}
\hline
 & a & b (GeV$^{-2}$) \\
\hline
set A & 0.3 & 0.15 \\
set B & 0.5 & 0.9 \\
\hline
\end{tabular}
\label{tab:set}
\end{table}

\begin{figure}
\centering
\includegraphics[width=0.45\textwidth]{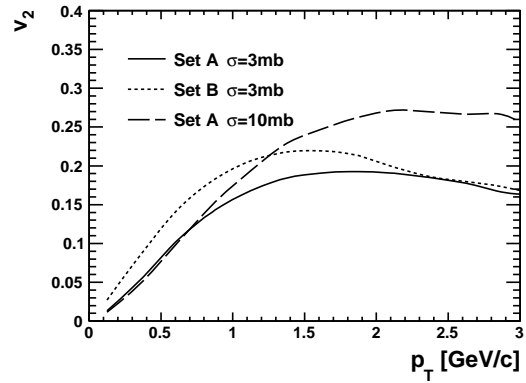}
\caption{Differential $v_2$ on $p_T$ of mid-rapidity charged pion in Pb-Pb collisions at $\snn=2.76$ TeV for 
centrality 30-40\% with different input parameters. The particles are formed right
after coalescence procedure.}
\label{fig:v2_pion_compare}
\end{figure}

\begin{figure}[hbt]
\centering
\includegraphics[width=0.45\textwidth]{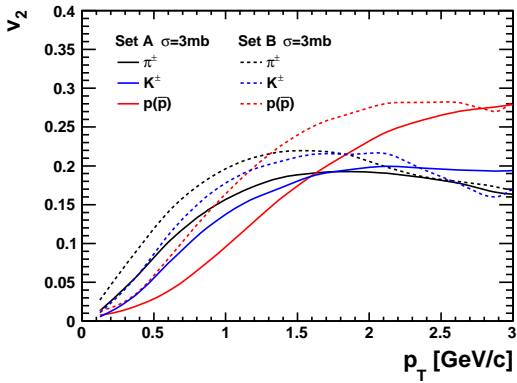}
\caption{(Color online)Differential $v_2$ on $p_T$ for selected particles at mid-rapidity
with different input parameters in Pb-Pb collisions at $\snn=2.76$ TeV for centrality 30-40\%
based on fixed parton rescattering cross section $\sigma=3$ mb. The particles are formed right
after coalescence procedure.}
\label{fig:v2_pikp_compare}
\end{figure}

We show in Fig.~\ref{fig:v2_pion_compare} the comparison of
different input parameters on the primordial charged pion $v_2$ as a function of
$p_T$. It can be found in this comparison that larger parton-parton cross
section leads to a much stronger $v_2$ in high $p_T$ region, while the low $p_T$
region $v_2$ is only slightly suppressed. On the other hand, for the fixed
$\sigma=3$ mb, input parameters with set B generate significantly larger $v_2$
in the low $p_T$ region, while both the two sets reach a similar $v_2$
for large $p_T$. We also make a comparison for the selected hadron $v_2$ in
Fig.~\ref{fig:v2_pikp_compare} based on different string tensions. Although the
overall flow is enhanced with set B parameters, the particle $v_2$ still
exhibits the mass ordering feature at low $p_T$ for both set A and set B.
It is pointed out in earlier studies that this ordering effect is mainly 
from kinematics in the quark coalescence process and later hadronic
rescatterings~\cite{Li:2016flp}. 
Thus, one would naturally expect it is independent of the initial state
conditions.

\begin{figure*}[htb]
\centering
\includegraphics[width=0.9\textwidth]{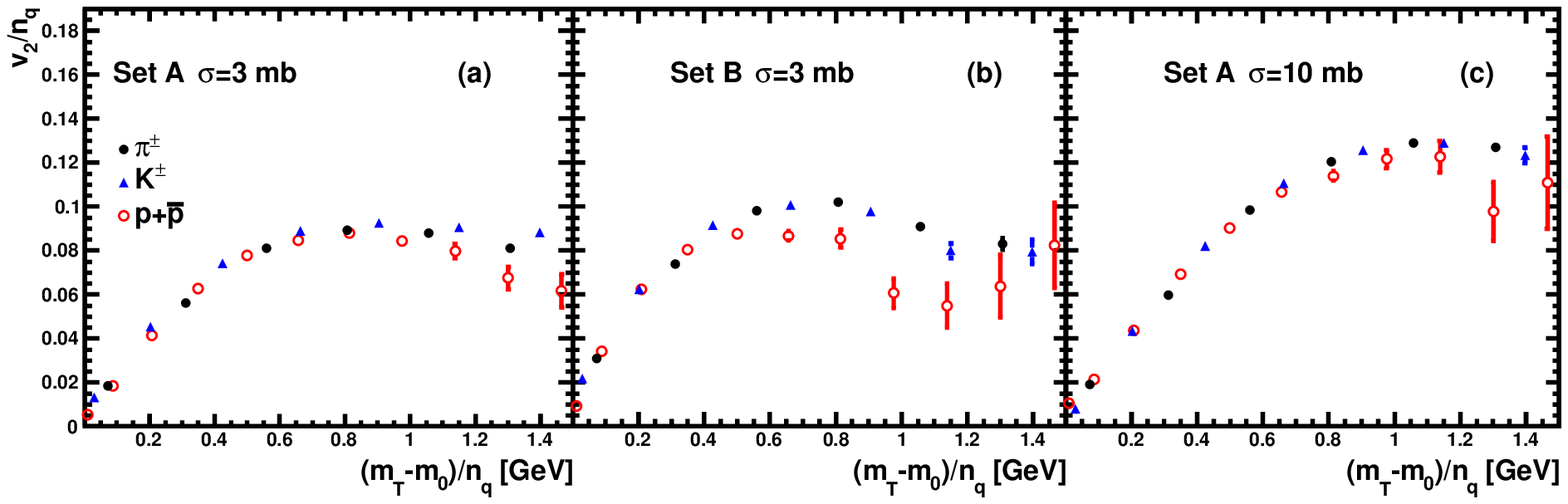}
\caption{(Color online) Number of constituent quark scaled $v_2$ for selected hadrons right after coalescence
procedure with parameters Set A $\sigma=3$ mb, Set B $\sigma=3$ mb, and
Set A $\sigma=10$ mb for Pb-Pb collisions at $\sqrt{s_{NN}}$=2.76 TeV with
the centrality 30-40\%. Charged pion, kaon and proton results
are labeled by the solid dots, triangles and open circles, respectively.}
\label{fig:ncq_before_compare}
\end{figure*}

At the intermediate $p_T$ region, the meson and baryon flow are supposed to
be amplified two/three-fold compared to that of the constituent quarks, leading to the number 
of constituent quark scaling behavior. This behavior can be well explained
in the quark coalescence mechanism, when the constituent quarks coalesce into
a meson or baryon in a collimated way.
%A naive quark coalescence process is expected to introduce a well
%established NCQ scaling of elliptic flow when parton phase space density is small.
We investigated the $v_2$ NCQ scaling behavior of some selected hadrons shown in
Fig.~\ref{fig:ncq_before_compare} based on a variety of input parameter sets for
the Pb-Pb collisions at $\snn=2.76$ TeV in centrality 30-40\%. It is found
that the NCQ scaling phenomena relies on the choice of the input parameters used
to produce the initial partons. The scaling behavior is evidently
violated with the set B input parameters while it still holds to a large extent
using the parameters of set A. By changing the parton-parton interaction cross
section from $\sigma=3$ mb to $\sigma=10$ mb, one can see the $v_2$ with set A
parameters still exhibits the NCQ scaling within the intermediate $p_T$ range of
$0.3<(m_{T}-m_{0})/n_{q}<1$ GeV.

To quantify the size of the violation to NCQ scaling, 
we defined the following quantity
\begin{equation}
\chi=\sqrt{\Sigma_{\pi,K}(\frac{v_{2}^{\pi,K}/n_{q}^{\pi,K}-v_{2}^{p}/n_{q}^{p}}{v_{2}^{p}/n_{q}^{p}})^{2}},
\label{eq:chi}
\end{equation}
where the $v_{2}$ are obtained for the integrated flow in the range
$0.3<(m_{T}-m_{0})/n_{q}<1$ GeV. Larger $\chi$ indicates stronger violations. It
is found that the magnitude of the violation to the NCQ scaling shown in 30-40\%
set A and 30-40\% set B can be
quantified by $\chi$ as: 0.06, 0.14, respectively. Smaller $\chi$
exists in the set A parameter results is consistent with our
expectations.

\begin{figure}
\centering
\includegraphics[width=0.45\textwidth]{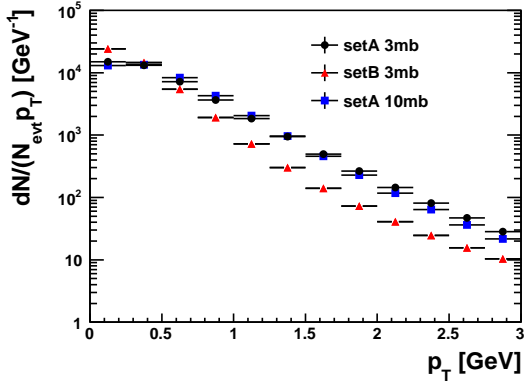}
\caption{$p_T$ distribution of partons at freeze out in Pb-Pb collisions at $\snn=2.76$ TeV for 
centrality 30-40\% with different input parameters and parton rescattering cross sections.}
\label{fig:parton_pt_compare}
\end{figure}

As it is known that the change of fragmentation parameter set will modify the initial
quark momentum and multiplicity distribution from string melting, we explored the parton
number distribution dependent on the transverse momentum at the time of coalescing to
final state hadrons in Fig.~\ref{fig:parton_pt_compare}. It is observed in this plot the
partons from initial conditions generated with set B parameter are much softer than that
with set A. With the same initial geometry configuration, the partons generated with set B
stay in the overlap region effectively longer and leading to a higher parton-parton
interaction rate, which breaks the initial scaling pattern. On the other hand, the change
in parton rescattering cross section from $\sigma=$ 3 mb to 10 mb is not casting much
impact on the freeze out parton momentum distribution especially in the $p_T$ range
relevant for NCQ scaling. Thus, one may not see any variation to the NCQ scaling from
parton rescattering.

It can be understood in this
comparison that the scaling behavior is related to the relative size of identified
particle $v_2$, thus decoupled from the global $v_2$ strength governed by the
parton-parton interaction cross section. On the other hand, the initial parton
conditions generated by different fragmentation parameters $a$, $b$ may play an
important role in determining the collimation of constituent quarks during the
coalescence of hadrons, which in the end modifies the $v_2$ NCQ scaling behavior.
In the following studies, we will rely on the input parameters of set A along with 
$\sigma=$ 3 mb to systematically explore the stability of $v_2$ NCQ scaling.

\subsection{Centrality dependence of NCQ scaling}
\label{subsec:results_sys}
One can observe the NCQ scaling with the input parameters of set A in AMPT model
for Pb-Pb semi-central collisions at the LHC energy, it will be interesting to
investigate the robustness of this scaling behavior with different collision centralities.
For this purpose, we will study the centrality dependence of the NCQ scaling in
Pb-Pb collisions at $\snn=$2.76 TeV.

\begin{figure}[hbt]
\centering
\includegraphics[width=0.52\textwidth]{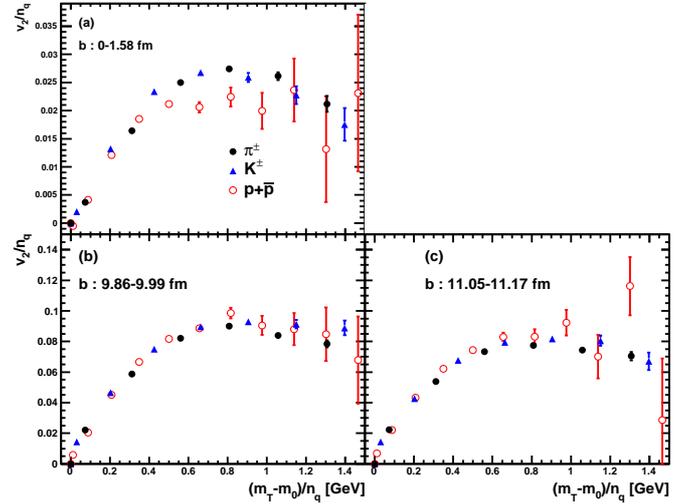}
\caption{(Color online) Number of constituent quark scaled $v_2$ for selected hadrons right after coalescence
procedure with parameters Set A and $\sigma=3$ mb for the centrality 0-1\% (b: 0-1.58 fm), 39-40\% (b: 9.86-9.99 fm) and 49-50\% (b: 11.05-11.17 fm) 
in Pb-Pb collisions at $\sqrt{s_{NN}}$=2.76 TeV. Charged pion, kaon and proton results
are labeled by the solid dots, triangles and open circles, respectively. 
}
\label{fig:ncq_before_centrality}
\end{figure}

Fig.~\ref{fig:ncq_before_centrality} shows the NCQ scaling behavior of some
selected hadrons within $|\eta|<0.8$ for the centrality of  0-1\% (impact parameter b: 0-1.58 fm), 
39-40\% (b: 9.86-9.99 fm) and 49-50\% (b: 11.05-11.17 fm). 
The centrality window has been fixed to a narrow range to remove the
centrality variation effects. One can observe in the most central bin 0-1\%, the
$n_q$ scaled $v_2$ for proton is systematically smaller than that for pion and
kaon, while in peripheral collisions the NCQ scaling is well preserved.
We have performed a $\chi$ calculation with the published ALICE data~\cite{Abelev:2014pua}
and find similar conclusion that $\chi=0.29$ for centrality 0-5\% and $\chi=0.17$ for
centrality 30-40\%. 

\begin{figure}
\centering
\includegraphics[width=0.45\textwidth]{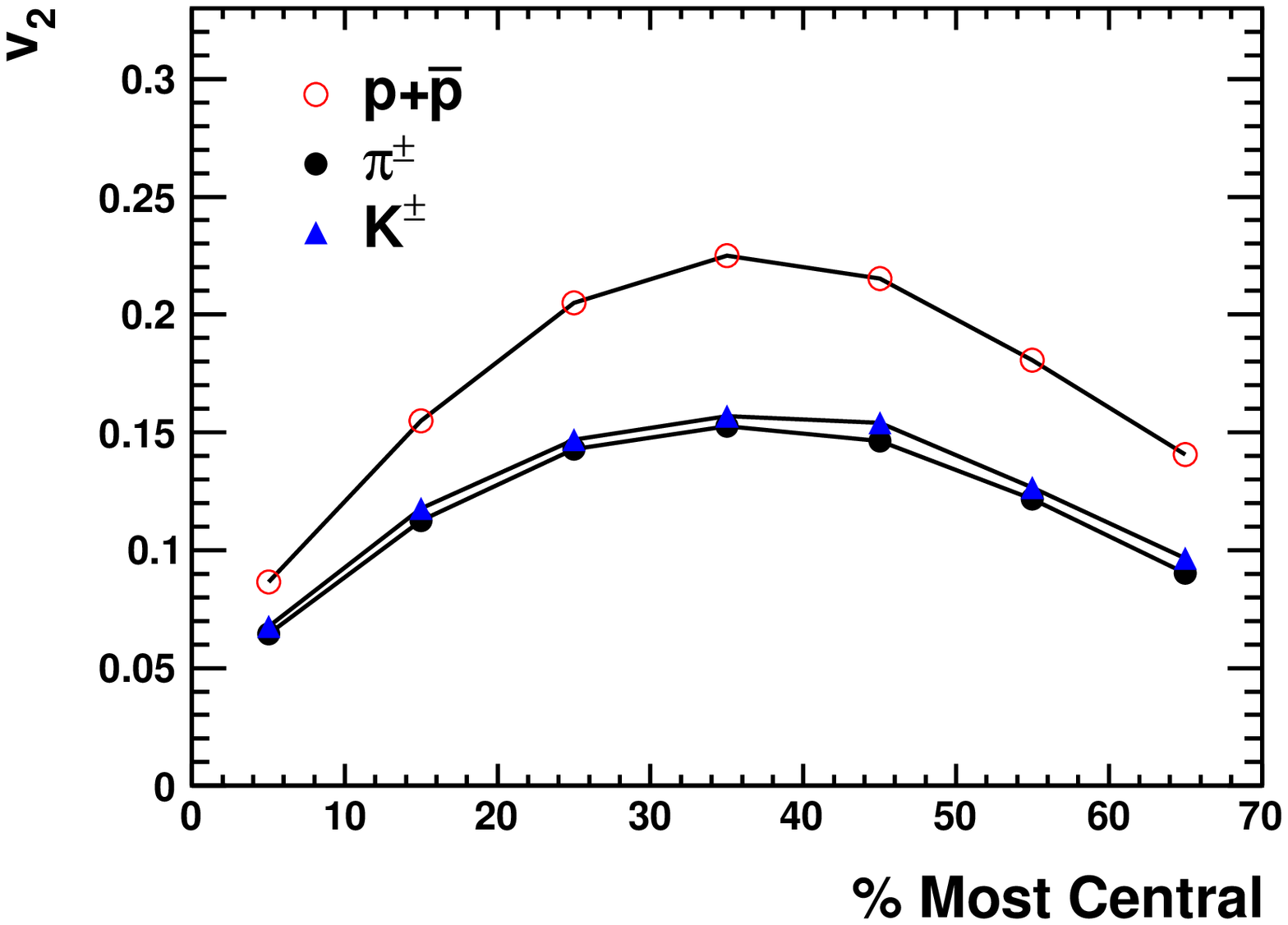}
\caption{(Color online) Integrated flow within $0.3<(m_{T}-m_{0})/n_{q}<1$ GeV for pion, kaon and
proton varying with the centrality.
}
\label{fig:ncq_before_integral_centrality}
\end{figure}

We integrated the $p_T$ differential flow over the corresponding range in
$0.3<(m_{T}-m_{0})/n_{q}<1$ GeV in several centrality bins between 0-70\% with
the result shown in Fig.~\ref{fig:ncq_before_integral_centrality}. The
integrated flow for all three particle species rises from central to semi-central
collisions and then drops after reaching the maximum in semi-central collisions.

\begin{figure}
\centering
\includegraphics[width=0.45\textwidth]{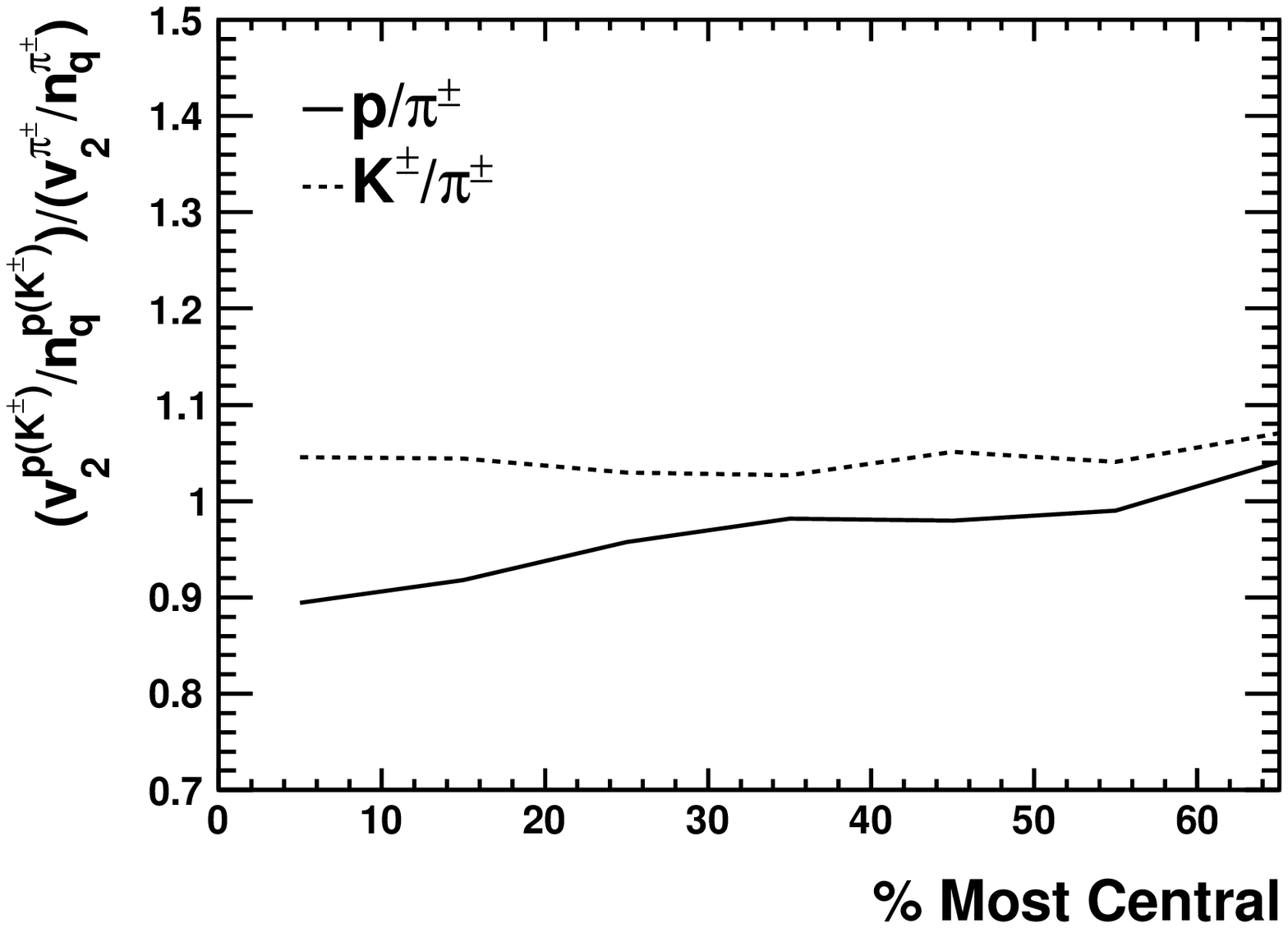}
\caption{Ratio of integrated flow within $0.3<(m_{T}-m_{0})/n_{q}<1$ GeV for proton over pion and 
kaon over pion varying with the centrality.
}
\label{fig:ncq_before_ratio_integral_centrality}
\end{figure}

There is also a trend shown in
Fig.~\ref{fig:ncq_before_ratio_integral_centrality} that the $n_q$ scaled
integrated $v_2$ between proton and pion grows monotonously from central to
peripheral collisions, approaching unity which satisfies the NCQ scaling within
the centrality range 30-60\%. On the other hand, the ratio between $K$ and $\pi$
is flatly distributed around unity for all centralities. 

To sum up, we observe that the NCQ scaling of $v_2$ is violated in
Pb-Pb central collisions, while it is recovered during the transition to the peripheral
collisions.

\subsection{Impact of hadronic interactions for identified particle elliptic flow}
\label{subsec:results_hadeff}

One postulation for the violation to the flow NCQ scaling at LHC energy is the
distortion of $v_{2}$ developed at partonic stage by later hadronic
interactions. We explored the impacts of possible different hadronic
interactions (including resonance decay and hadron rescattering in hadronic
evolution stage) on the $v_2$ of some selected hadrons. We switched on the
contribution from resonance decay and hadronic rescattering independently and
compared the impacts of these hadronic effects with respect to the flow
developed for the primordial particles mainly from quark evolution phase.
The hadron rescattering strength is mediated by maximum allowed hadronic 
rescattering time $t_{max}$ in AMPT.

\begin{figure*}
\centering
\includegraphics[width=0.9\textwidth]{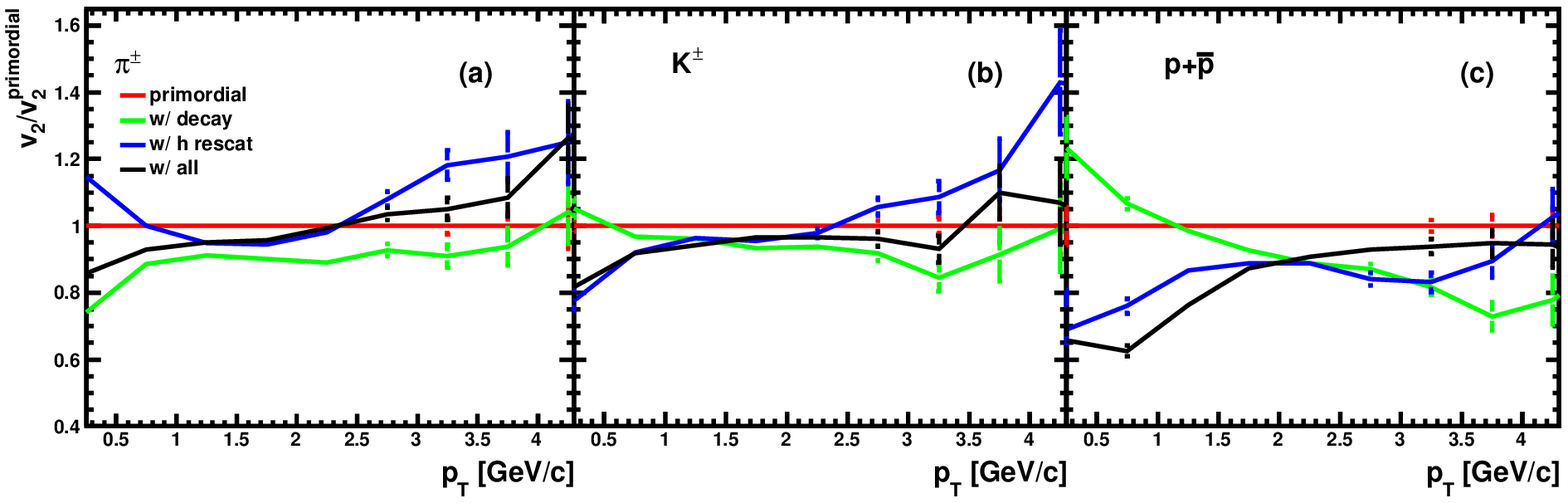}
\caption{(Color online) Ratio of $v_2$ with different hadronic effects divided
by that right after coalescence for pions (a), kaons (b) and
protons (c). The elliptic flow is calculated for Pb-Pb collisions at
$\sqrt{s_{NN}}=$ 2.76 TeV with the centrality of 30-40\% using parameters set A
and $\sigma=$ 3 mb. $t_{max}=$ 30 fm/c for the allowed hadron evolution time.}
\label{fig:v2ratio_piKp_peripheral_hadron_effect}
\end{figure*}

It is shown in Fig.~\ref{fig:v2ratio_piKp_peripheral_hadron_effect} that the
comparison of the $p_T$ differential $v_2$ with different hadronic effects
divided by that for the primordial particles formed right after coalescence.
These comparisons use the default hadronic rescattering time $t_{max}=30$ fm/$c$.
Hadronic rescattering seems to destruct the azimuthal anisotropy formed during the
quark stage at low $p_T$ for protons and kaons, while resonance decay enlarges low $p_T$ $v_2$
and reduces high $p_T$ $v_2$ for these two particle species. On the other hand, the pion
$v_2$ is following the opposite trend. A kink like behavior is observed
for pion hadron rescattering dominated flow ratio. If we put on the resonance
decay along with hadron rescattering step by step, a suppression is observed
for the intermediate $p_T$ flow of all three particles and for proton this
effect is much stronger. Indicated by this comparison, the distortion of elliptic flow from
hadronic evolution is an important source for the violation of NCQ scaling.
%resonance decay plays a significant role in the formation of $v_2$ for pions while
%hadron rescattering is important for kaons, especially when $p_T$ is below 1
%GeV. As to the protons, hadron rescattering brings in a strong drop to the
%initial $v_2$ for $p_T$ less than 2 GeV. However, both hadron rescattering and
%resonance decay becomes equally important in the formation of proton $v_2$ for
%$p_T$ greater than 2 GeV. The difference of the impact from

\begin{figure}
\includegraphics[width=0.45\textwidth]{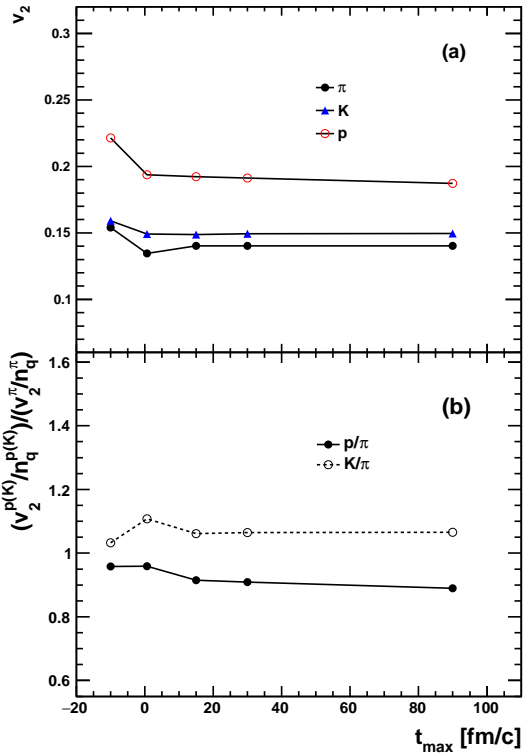}
\caption{(Color online) Integrated flow for selected particles within
$0.3<(m_{T}-m_{0})/n_{q}<1$ GeV for pion, kaon and proton (a) along with the
integrated flow ratio of proton over pion and kaon over pion (b) varying with $t_{max}$. 
The elliptic flow is calculated for Pb-Pb collisions at $\sqrt{s_{NN}}=$ 2.76 TeV 
with the centrality of 30-40\% using parameters set A and $\sigma=$ 3 mb.
}
\label{fig:integral_afterART}
\end{figure}

We studied the hadron interaction strength dependence of the identified particle
flow integrated over $0.3<(m_{T}-m_{0})/n_{q}<1$ GeV in
Fig.~\ref{fig:integral_afterART} by varying $t_{max}$ parameter in AMPT. Larger
$t_{max}$ suggests stronger hadronic interaction contributions. The
notation $t_{max}=-$10 fm/$c$ denotes the flow of primordial particles formed right after
coalescence. $v_2$ with $t_{max}=$ 0.3 fm/$c$ effectively turns off the hadron
rescattering but includes the resonance decay. As can be seen, resonance decay
destroys the collectivity for all three particles in the intermediate $p_T$
range, comparing $t_{max}=-$10 fm/$c$ to $t_{max}=$ 0.3 fm/$c$. The integrated$v_2$ 
decreases with $t_{max}$ in general, although the dependence is
very weak when $t_{max}$ is greater than 15 fm/$c$. The difference between proton
over pion $v_2$ ratio and unity becomes enlarged with rising $t_{max}$ suggests
that larger violation to NCQ scaling is generated due to stronger hadronic
interactions.

In order to understand the dependence of the hadron evolution impact on
the violation to NCQ scaling, we listed the values of $\chi$
calculated at different time scales in Tab.~\ref{tab:chi2} based on Eq.~\ref{eq:chi}. 
It is observed that the magnitude of violation to NCQ scaling before hadronice
interaction stage is relatively small as low as 0.06. In case the resonance
decay is included, the violation degree rapidly grows to 0.16. The impact
of hadron rescattering is dependent on the allowed maximum time length of hadron
evolution.

\begin{table}[hbt]
\centering
\caption{$\chi$ at different time scale}
\begin{tabular}{p{0.2cm}|p{1.35cm}|p{1.28cm}|p{2cm}|p{2cm}}
%\toprule
\hline
 & primordial & w/ decay & $t_{max}$=15 fm/$c$ & $t_{max}$=90 fm/$c$ \\
\hline
$\chi$ & 0.06 & 0.16 & 0.19 & 0.24\\
\hline
\end{tabular}
\label{tab:chi2}
\end{table}

\subsection{$v_2$ NCQ scaling at $\sqrt{s_{NN}}$=5.02 TeV}
\label{subsec:results_prediction}
The particle yield and azimuthal anisotropy size are found to slightly increase
with the collision energy from $\sqrt{s_{NN}}$ = 2.76 TeV to $\sqrt{s_{NN}}$ = 5.02
TeV. It is then expected to see little variation on the NCQ scaling behavior of
the identified particle elliptic flow with the top Pb-Pb collision energy at the
LHC. We will explore the $v_2$ NCQ scaling within AMPT model at
$\sqrt{s_{NN}}$=5.02 TeV based on the input parameter set A with $\sigma=$ 3 mb,
These parameters have been extensively used in the predictions on various observables in Pb-Pb
collisions at $\sqrt{s_{NN}}$=5.02 TeV~\cite{Ma:2016fve}.

\begin{figure}
\centering
\includegraphics[width=0.52\textwidth]{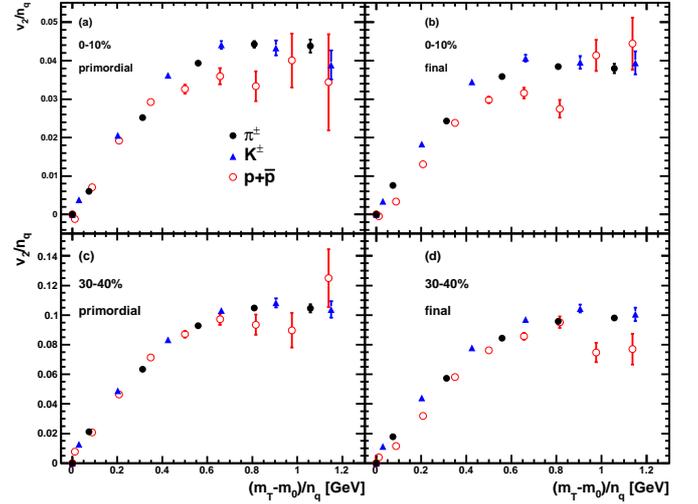}
\caption{(Color online) $v_2$ NCQ scaling at $\sqrt{s_{NN}}=$ 5.02 TeV with the
centrality 0-10\% and 30-40\% using parameters set A, $\sigma=$ 3 mb before
(Left column) and after (Right column) hadronic interactions.
}
\label{fig:502_compare}
\end{figure}

It can be found in Fig.~\ref{fig:502_compare} that the $n_q$ scaling works to a
large extent before hadronic evolution takes place in the semi-central collisions with
centrality 30-40\%. The centrality dependence and hadronic interaction impact
are similar to what has been uncovered in the studies for $\sqrt{s_{NN}}$=2.76
TeV in this work. Hadronic interactions are also important to account for the
violation of NCQ scaling at this energy scale. 

\section{Summary}
In this work, we have systematically investigated the NCQ scaling of elliptic flow at the
LHC energy using the AMPT model. We find the NCQ scaling exists with the initial
conditions generated in the semi-central collisions based on the Lund string parameter set
carrying larger string tension if the hadronic interaction is off. The larger number of
initial partons from the conditions generated with fragmentation parameter carrying
smaller string tension or in central collisions could lead to more scatterings and violate
NCQ scaling in the current AMPT framework. It is shown in these comparisons that the NCQ scaling
structure not only depends on the hadronization procedure but also relies on the parton
dynamics at the initial stage before evolution.

%Two sets of parameters used to
%generate the initial conditions in AMPT have been compared in our study.
%With the effective temperature  $T_{<p_{T}>}=\frac{4<p_{T}>}{3}$ used in the work of~\cite{Lin:2014tya},
%one can obtain the $T_{<p_{T}>}$ 199 MeV for set A and 136 MeV for set B
%parameters, respectively.

The impact of the hadronic interaction on the $v_2$ of selected hadrons has
been studied in details. Resonance decay and hadron rescattering modify the
$v_2$ magnitude differently for different particle species. A sizable distortion
to the NCQ scaling arises due to the effects introduced by the
hadronic interactions. Our quantitative analysis shows the contribution
from resonance decay and hadron rescattering are both important
in the development of the violation to NCQ scaling.

Besides, it must be aware that the parton coalescence mechanism in the current
AMPT model is performed in the coordinate space of constituent quarks. There is
a chance that the momentum collimation is not well preserved during the
coalescence procedure, which might lead to a violation to the exact NCQ scaling.
It can be of great interest to test the scaling behavior from a coalescence
procedure implemented with collimated quarks in both momentum and
coordinate space in the future.

%\vspace{4\baselineskip}

\begin{acknowledgement}
We thank Zi-Wei Lin, Guo-Liang Ma and Jun Xu for helpful discussions. This work was supported by NSFC (11475068 and 11605070), China Postdoctoral Science Foundation (2016M590703), the National Key Research and Development Program of China under Grant No.2016YFE0100900. LZ acknowledges support by the programme of Introducing Talents of Discipline to Universities (B08033). 
\end{acknowledgement}

\bibliography{reference}
\bibliographystyle{bibstyle/epjstyle}

\end{document}